\documentclass[a4paper,12pt]{article}
\usepackage{amsmath,amsthm,amssymb}
\usepackage[dvipdfm]{graphicx}

\title{On the Independent Crossing Approximation
in Three Flavor Neutrino Oscillations}
\author{Kazunari Yamamoto\\ Department of Physics, Tokyo Metropolitan University,\\ Minami--Osawa, Hachioji, Tokyo 
192--0397, Japan}
\date{\today}
\begin{document}
\maketitle

\begin{abstract}
In dense matter like the Supernovae and at high energy, neutrinos
have two points where the nonadiabatic transitions could occur. With
the present values of the oscillation parameters in the standard three
flavor scenario, two nonadiabatic transitions can be regarded as
independent, i.e., two can be treated by the two flavor formalism
separately. However, in the presence of new physics,
it is not clear in general whether such independence holds.
We examine this question by assuming
hypothetical range of the neutrino energy and by varying the mixing angles
and the mass squared differences in the standard case. We found the
cases where Independent Crossing Approximation breaks down for some unrealistic
range of $\Delta m^2_{31}$ and $\theta_{13}$. We also discuss the
criterion which gives such independence.
\end{abstract}

\newpage
\section{Introduction}
We now understand the solar neutrino problem~\cite{pdg:2008}, i.e., we now know that the deficit of the electron neutrino flux from the Sun can be accounted for approximately as the two 
flavor neutrino oscillations in matter. From the precise experiments, we know the mixing angle of the 
neutrino oscillations is large. It is called Large Mixing Angle solution. Historically, the so-called Small Mixing Angle solution has also been discussed~\cite{Bahcall:2000kh}, because it gave good fit to the solar neutrino data in the past. In this case, 
the nonadiabatic transitions, which are jumping between the different neutrino energy levels in matter, would become important.   On the other hand, we 
also understand now the atmospheric neutrino problem~\cite{pdg:2008}, i.e., the deficit of the muon neutrino flux from the cosmic ray can be explained approximately as 
the two flavor neutrino oscillations. We also know the oscillation channel of the atmospheric neutrino differ from the 
solar neutrino. These lead us to conclude that the neutrino oscillation is described by three flavor neutrino oscillations. 

In the case of the environments with high density, such as supernovae, it is important to consider the neutrino 
oscillations with the three flavor exactly, since there can be two energy level crossings. 
Furthermore, if nonadiabatic transitions occur at each level crossing at all, then we need to take them into account. In 
this case, there could be two nonadiabatic transitions~\cite{Dighe:1999bi}. The 
three flavor neutrino oscillations with two nonadiabatic transitions are very complicated in general. Although little is 
known about the exact solution of this very complicated problem, for instance,
Ref.~\cite{Kuo:1989qe} assumes the Independent Crossing Approximation (ICA), which regards two nonadiabatic transitions as 
independent.
In this approximation, we can regard each 
nonadiabatic transition locally as that between the two levels, and then we can get the total nonadiabatic transition just by 
multipling each probability of the nonadiabatic transition~\cite{Kuo:1989qe}: 

$$\hat{P}_c = \hat{P}_\mathrm{Lower} \times \hat{P}_\mathrm{Higher}$$. 

The reason that ICA has been assumed is (i) because the two energy
ranges which give a resonance do not overlap for the solar and
atmospheric neutrino oscillation parameters and (ii) because the
energy of the supernovae neutrino is so low and the solar neutrino
mixing angle is so large that we would not have nonadiabatic
transitions at both of the energy level crossings.

However, if we have new physics beyond the standard model, then the
standard matter effect may be modified~\cite{Wolfenstein:1977ue,Guzzo:1991hi,Roulet:1991sm}
and it becomes unclear whether these reasonings always hold.
The purpose of this paper is to examine whether ICA holds or
not in the three flavor neutrino oscillations, and to discuss under
which condition ICA breaks down.  For simplicity we take only the
three flavor case with the standard matter effect as an example and
\textit{we will assume hypothetical neutrino energy and hypothetical
values of the oscillation parameters} to have nonadiabatic transitions
at both energy level crossings.

From numerical calculations, we found that ICA breaks down for
some range of $\Delta m^2_{31}$ and $\theta_{13}$.  We found that ICA
breaks down when the ratio $\Delta m^2_{31}/\Delta m^2_{21}$ is of
order one and the mixing angle $\theta_{13}$ is large.  In the
realistic range of the oscillation parameters, since $\Delta
m^2_{31}/\Delta m^2_{21}$ is much larger than one and the mixing angle
$\theta_{13}$ is small, the two neutrino energy ranges for the
resonance never overlap.
This is the reason why the treatment in Ref.~\cite{Kuo:1989qe} is
regarded as correct. However, it turns out that it is not the
correct criterion because we find that two level crossings become
independent even if the two resonances overlap.  Instead of using the
notion of the overlapping resonances,
we will introduce the new parameters which gives the criterion
of ICA.  With this criterion of ICA, we interpret the breaking
of ICA as the sign of a large contribution of the extra off--diagonal
coefficients.  Our numerical calculations confirm this analytic
treatment very well.

Although we discuss the simplest case only with the neutrino--electron
interaction in this paper, we can apply our treatment to any other
cases even if it is not known whether the nonadiabatic transitions are
independent or not. For example, even if the effective ratio $(\Delta
\widetilde{m}^2_{31}/\Delta \widetilde{m}^2_{21})$ in the
presence of new physics is small, or the effective
$\widetilde{\theta}_{13}$ is large, we can discuss independence of the
nonadiabatic transitions.

\section{Neutrino oscillations with two flavor in matter}
In this section, we review the standard treatment of the neutrino oscillations with two flavor. 
\subsection{Adiabatic transitions}
The positive energy part of the Dirac equation for the flavor eigenstates $\nu_\alpha(t)
\,(\alpha=e, \mu)$ propagating at time $t$ in 
matter is given by
\begin{eqnarray} \label{eq:2flavor}
i\frac{d}{dt}\nu_\alpha(t) &=& H\,\nu_\alpha(t), 
\label{eq:dirac}
\end{eqnarray}
where $H$ is given by 
\begin{eqnarray}
H &=& U\left(
			   \begin{array}{cc}
			   0 & 0 \\
			   0 & \Delta
			   \end{array}
		     \right)
			    U^{-1} + \left(
			   		   \begin{array}{cc}
			   		   A & 0 \\
			   		   0 & 0
			   		   \end{array}
		     		     \right),
\end{eqnarray}
here $\Delta \equiv \Delta m^2/2E$, $A \equiv \sqrt{2}G_\mathrm{F}n_e(R)$ is the extra potential in medium at a distance $R=ct$ from the initial point, and $U$ is 
the $2 \times 2$ MNS matrix~\cite{Maki:1962mu}. $H$ can be diagonalized as 
\begin{eqnarray}
H = \exp(i\sigma_2\widetilde{\theta}) \mbox{\rm diag}(\widetilde{E}_1, \widetilde{E}_2) 
\exp(-i\sigma_2\widetilde{\theta}). 
\end{eqnarray}
where $\sigma_2$ is the Pauli matrix and the effective mixing angle $\widetilde{\theta}$ is given by 
\begin{equation} \label{eq:appamp}
\sin^22\widetilde{\theta} = \frac{(\Delta\sin2\theta)^2}{(\Delta\cos2\theta - A)^2 + (\Delta\sin2\theta)^2} \, . 
\end{equation}
$\sin^22\widetilde{\theta}$ becomes the maximum at the point $A(R_{\mathrm{res}}) = \Delta\cos2\theta$, which is called 
the Resonance point. 

In the adiabatic case, the positive energy part of the Dirac equation for the mass eigenstates can be written as
\begin{equation}\label{eq:adiabatic}
i\frac{d}{dt}\left(
		  \begin{array}{c}
		  \widetilde{\nu}_1 \\
	          \widetilde{\nu}_2
	          \end{array}
		\right) = \left(
     				\begin{array}{cc}
     				\widetilde{E}_1 & 0 \\
     				0 & \widetilde{E}_2
     				\end{array}
   			  \right) \left(
	    				\begin{array}{c}
	    				\widetilde{\nu}_1 \\
            				\widetilde{\nu}_2
            				\end{array}
	  			  \right), 
\end{equation}
where $\widetilde{\nu}_j\equiv\exp(-i\sigma_2\widetilde{\theta})\,\nu_\alpha$
\,($j=1, 2$).
Integrating Eq.(\ref{eq:adiabatic}), we obtain the survival probability of the electron neutrino 
\begin{eqnarray} \label{eq:pvm}
P(\nu_e \rightarrow \nu_e ;R)
= \cos^2\left(\widetilde{\theta}(R) - \widetilde{\theta}(0)\right)
    - \sin\frac{\widetilde{\theta}(R)}{2}\sin\frac{\widetilde{\theta}(0)}{2}\sin^2\int^R_0\frac{\Delta\widetilde{E}}{2}dr,
\end{eqnarray}
where the difference of the eigenvalues is 
\begin{equation}
\Delta\widetilde{E} \equiv \widetilde{E}_2 - \widetilde{E}_1
= \sqrt{(\Delta\cos2\theta - A)^2 + (\Delta\sin2\theta)^2}. 
\end{equation}
In the case of the solar neutrino, since $\int^R_0\Delta\widetilde{E}dr \gg 1$ and $|A(0)/\Delta| \gg 1$, we get 
\begin{equation} \label{eq:pvsol}
P(\nu_e \rightarrow \nu_e) = \sin^2\theta. 
\end{equation}

\subsection{Nonadiabatic transition}
Let us now discuss Eq.(\ref{eq:dirac}) in the nonadiabatic case, i.e., in the case where we cannot ignore the variation of 
the effective mixing angle. In this case, we have 
\begin{eqnarray} \label{eq:off}
i\frac{d}{dt}\left(
		  \begin{array}{c}
		  \widetilde{\nu}_1 \\
	          \widetilde{\nu}_2
	          \end{array}
		\right)
= \left(
     \begin{array}{cc}
     \widetilde{E}_1 &  -i\dot{\widetilde{\theta}} \\
      i\dot{\widetilde{\theta}} & \widetilde{E}_2
     \end{array}
   \right)
                     	    \left(
	    		      \begin{array}{c}
	    		      \widetilde{\nu}_1 \\
            		      \widetilde{\nu}_2
            		      \end{array}
	  		    \right).
\end{eqnarray}
The off--diagonal elements in Eq.(\ref{eq:off}) stand for an effect in which neutrino jumps the energy gap between 
$\widetilde{E}_1$ and $\widetilde{E}_2$, and such an effect becomes nonnegligible when the following 
the adiabatic condition breaks down: 
\begin{equation} \label{eq:ac}
\gamma \equiv \left|\frac{\Delta m^2}{2E}\frac{\sin^22\theta}{\cos2\theta \, (\dot{n_e}/n_e)_\mathrm{res}}\right| \gg 1,
\end{equation}
which can be derived by comparing the diagonal and off--diagonal elements at the resonance point.\footnote{Here the 
resonance point differs from the Point of Maximal Violation of Adiabaticity~\cite{Kachelriess:2001fs,Friedland:2001xk}.} 

Zener~\cite{Zener:1932ws} derived the jumping probability of spins in a linear magnetic field. In order for Zener's method 
to apply for neutrino, the density profile of electron has to be approximately linear. In that case, using the adiabatic 
condition Eq.(\ref{eq:ac}), the jumping probability is given by 
\begin{subequations} \label{eq:zkp}
\begin{equation} \label{eq:zener}
P_{\mathrm{Zener}} = \exp\left(-\frac{\pi}{2} \, \gamma\right). 
\end{equation}

Kuo and Pantaleone~\cite{Kuo:1988pn} showed that the differential equation for the exponential profile can be solved 
exactly. The analytical expression for the jumping probability is given by 
\begin{equation} \label{eq:exact}
P_{\mathrm{exact}} = \frac{\exp\left[\displaystyle -\frac{\pi}{2}\gamma(1 - \tan^2\theta)\right]
		- \exp\left[\displaystyle -\frac{\pi}{2}\gamma\left(\frac{1 - \tan^2\theta}{\sin^2\theta}\right)\right]}
	     {1 - \exp\left[\displaystyle -\frac{\pi}{2}\gamma\left(\frac{1 - \tan^2\theta}{\sin^2\theta}\right)\right]}.
\end{equation}
\end{subequations}
Using the jumping probability $P_c$ , we can write down a simple expression for the survival probability of the electron 
neutrino. Again by taking the limits $\int^R_0\Delta\widetilde{E}dr \gg 1$ and $|A(0)/\Delta| \gg 1$, we obtain 
\begin{eqnarray} \label{eq:2f}
P(\nu_e \rightarrow \nu_e) &=& \left(
				   \begin{array}{cc}
				   \cos^2\theta & \sin^2\theta
				   \end{array}
			     \right) \left(
					   \begin{array}{cc}
					   1 - P_c & P_c \\
					   P_c & 1 - P_c 
					   \end{array}
	  			     \right) \left(
						   \begin{array}{c}
						   0 \\
						   1
						   \end{array}
					     \right) \nonumber \\
			   &=& \sin^2\theta + P_c\cos2\theta. 
\end{eqnarray}

The formulae of Eqs.(\ref{eq:zkp}) for the jumping probability differ
when the mixing angle is large. In this case, especially for the
extreme nonadiabatic transition, i.e., $\gamma \rightarrow 0$,
the jumping probability $P_c$ reaches the maximum
value, i.e., complete conversion, in the
Zener's solution, but it does not in the exact solution.

\section{Neutrino oscillations with three flavor in matter}
The positive energy part of the Dirac equation with three flavor in matter is given by 
\begin{subequations} \label{eq:3flavors}
\begin{eqnarray} \label{eq:3flavor}
i\frac{d}{dt}\nu_\alpha(t) &=& H(t)\nu_\alpha(t), 
\end{eqnarray}
where $\alpha=e, \mu, \tau$, the Hamiltonian is 
\begin{eqnarray} \label{eq:3flavor2}
H(t) &=& U\left(
			   \begin{array}{ccc}
			   0 & 0 & 0\\
			   0 & \Delta_{21} & 0 \\
                           0 & 0 & \Delta_{31}
			   \end{array}
		     \right)
			    U^{-1} + \left(
			   		   \begin{array}{ccc}
			   		   A(t) & 0 & 0 \\
			   		   0 & 0 & 0 \\
                                           0 & 0 & 0
			   		   \end{array}
		     		     \right), 
\end{eqnarray}
\end{subequations}
and the standard parametrization~\cite{pdg:2008} for the $3 \times 3$ MNS matrix is 
\begin{eqnarray*}
U = \left(
  	    \begin{array}{ccc}
		c_{12} c_{13} & s_{12} c_{13} & s_{13} e^{- i \delta} \\
  		- s_{12} c_{23} - c_{12} s_{23} s_{13} e^{i \delta} & c_{12} c_{23} - s_{12} s_{23} s_{13} e^{i \delta} & 
                s_{23} c_{13} \\
  		s_{12} s_{23} - c_{12} c_{23} s_{13} e^{i \delta} & - c_{12} s_{23} - s_{12} c_{23} s_{13} e^{i \delta} & 
                c_{23} c_{13} \\
 	    \end{array} 
      \right).
\end{eqnarray*}
Here $s_{ij} \equiv \sin\theta_{ij}$, $c_{ij} \equiv \cos\theta_{ij}$, $\theta_{ij}$ are the mixing angles and $\delta$ is 
the CP phase as in the quark sector~\cite{Kobayashi:1973fv}. 

\subsection{ICA with two nonadiabatic transitions}
For the nonadiabatic transitions with three flavor, the neutrino can have two jumping points. When these two points are
approximately far apart, from the analogy with Eq.(\ref{eq:2f}), we get the survival probability of the electron 
neutrino~\cite{Kuo:1989qe}: 
\begin{eqnarray}
P(\nu_e \rightarrow \nu_e) &=& \left(
		 \begin{array}{ccc}
		 |U_{e1}|^2 & |U_{e2}|^2 & |U_{e3}|^2
		 \end{array}
    	   \right) \hat{P}_L\hat{P}_H \left(
	   				    \begin{array}{c}
				 	    0 \\
				 	    0 \\
                                 	    1
				 	    \end{array}
	   		   	      \right) \nonumber \\
&=& |U_{e1}|^2P_LP_H + |U_{e2}|^2P_H(1 - P_L) + |U_{e3}|^2(1 - P_H), 
\label{eq:3level}
\end{eqnarray}
where 
\begin{eqnarray*}
\hat{P}_L &=& \left(
		\begin{array}{ccc}
		1 - P_L & P_L & 0 \\
		P_L & 1 - P_L & 0 \\
		0 & 0 & 1
		\end{array}
	  \right) \\
\hat{P}_H &=& \left(
		\begin{array}{ccc}
                1 & 0 & 0 \\
		0 & 1 - P_H & P_H \\
		0 & P_H & 1 - P_H
		\end{array}
	  \right). 
\end{eqnarray*}
Here, $P_L$ and $P_H$ are the jumping probabilities at the lower and higher crossing point each other, and $U_{ei}$ are the 
elements of the three flavor MNS matrix. 

\subsection{The setting of our analysis}
What we would like to discuss here is whether the ICA in Eq.(\ref{eq:3level}) holds 
or not.
Let us 
start our study by assuming several situations. 

Throughout this paper, we will take the following reference values
for the oscillation parameters and the reference function for the
electron density $n_e(R)$ at distance $R$ from the initial point: 
\begin{subequations}
\begin{eqnarray}
&& \sin^22\theta_{12} = 0.87 \\
&& \sin^22\theta_{23} = 1.0 \\
&& \Delta m^2_{21} = 7.9 \times 10^{-5} \, \mathrm{eV}^2 \\
&& \delta = 0 \\
&& n_e(R) \simeq 50 \times n_{e\odot}(R) \quad \mathrm{cm}^{-3} \label{eq:density} \\
&& n_{e\odot}(R) \simeq 245N_A\exp(-10R/R_\odot) \quad \mathrm{cm}^{-3}, 
\end{eqnarray}
\end{subequations}
where $N_A$ is the Avogadro's number, $n_{e\odot}(R)$ stands for
the electron density in the solar standard model~\cite{Bahcall:2000nu},
and $R_\odot$ is the solar radius. 

A few remarks are in order.
(i) We refer to the standard value~\cite{pdg:2008} for the first
three oscillation  parameters.
(ii) We assume that the CP phase $\delta$ is equal to zero.  This is
because the transition probability of $\nu_e \rightarrow \nu_e$ (and
its antineutrino) did not depend on the CP phase
$\delta$~\cite{Yokomakura:2002av}.
(iii) We assume the electron density which is proportional to the one
in the Sun but is larger by a factor 50, in order to get the two
energy level crossing.\footnote{If $|A/\Delta_{31}| \gg 1$, then it follows
that $\widetilde{\nu}_e(0) \simeq \widetilde{\nu}_3(0)$} 
For the exponential density profile, we can get a simple value
$\dot{n}_e/n_e = \mathrm{const.}$ for Eq.(\ref{eq:ac}).

On the other hand,
In order to have nonadiabatic transitions at the two level crossings,
we will have to assume hypothetical values for the following parameters: 
\begin{eqnarray*}
&& \sin^22\theta_{31} \\
&& \Delta m^2_{31} \\
&& E,
\end{eqnarray*}
where we will assume $\Delta m^2_{31}>0$, i.e., we will assume
the so-called normal hierarchy, throughout this paper
for $\nu_e$ (instead of $\bar{\nu}_e$) to have two level crossings.

Furthermore, we adopt the expression Eq.(\ref{eq:exact}) for $P_L$ and
$P_H$ for the low (high). Here we take each of the adiabatic
conditions, $\gamma_L$ and $\gamma_H$ as two levels at each crossing
point. From the analogy with Eq.(\ref{eq:ac}) we have
\begin{subequations}
\begin{align} \label{eq:ac2}
\gamma_L &\equiv \left|\frac{\Delta_{21}\sin^22\theta_{12}}
				{\cos2\theta_{12} \, (\dot{n_e}/n_e)_\mathrm{res}}\right| \\ \nonumber \\
\gamma_H &\equiv \left|\frac{\Delta\sin^22\theta_{13}}
				{\cos2\theta_{13} \, (\dot{n_e}/n_e)_\mathrm{res}}\right|, 
\end{align}
\end{subequations}
where we have defined, 
\begin{eqnarray}
\Delta_{21} &\equiv& \Delta m^2_{21}/2E \nonumber\\
\Delta_{31} &\equiv& \Delta m^2_{31}/2E \nonumber\\
\Delta &\equiv& \Delta_{31} - \Delta_{21}\sin^2\theta_{12} \, . 
\label{eq:Delta}
\end{eqnarray}
In our study, we use the numerical calculation to check whether ICA holds or not.
We use the Runge--Kutta method to solve numerically the positive energy part of 
the Dirac equation for three flavor (see Eqs.(\ref{eq:3flavors})).

\begin{figure}[!ht]
\begin{center}
\includegraphics[width=8cm]{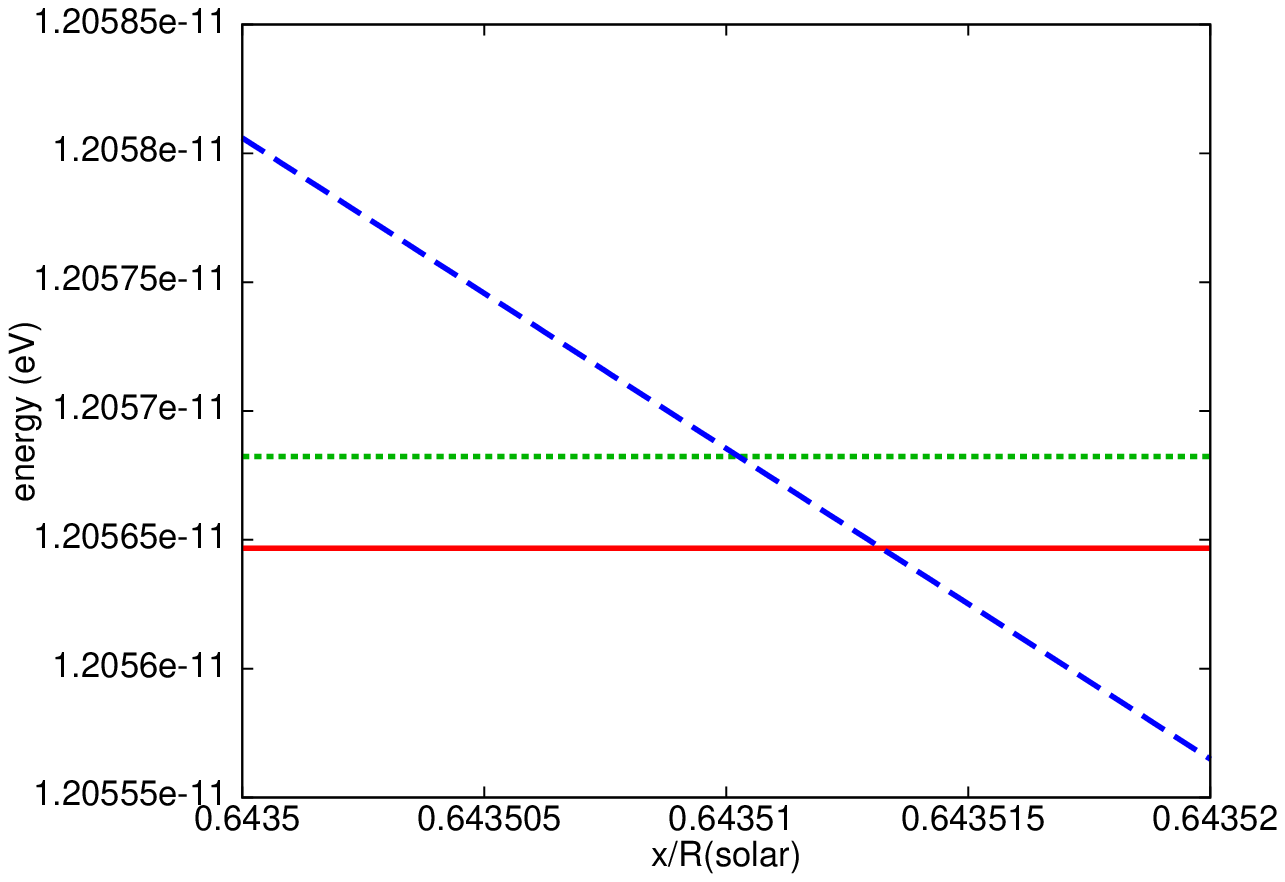}
\caption{The diagonal elements of the neutrino in matter. Two crossing points are close.} \label{fig:17}
\vspace*{2cm}
\includegraphics[width=11cm]{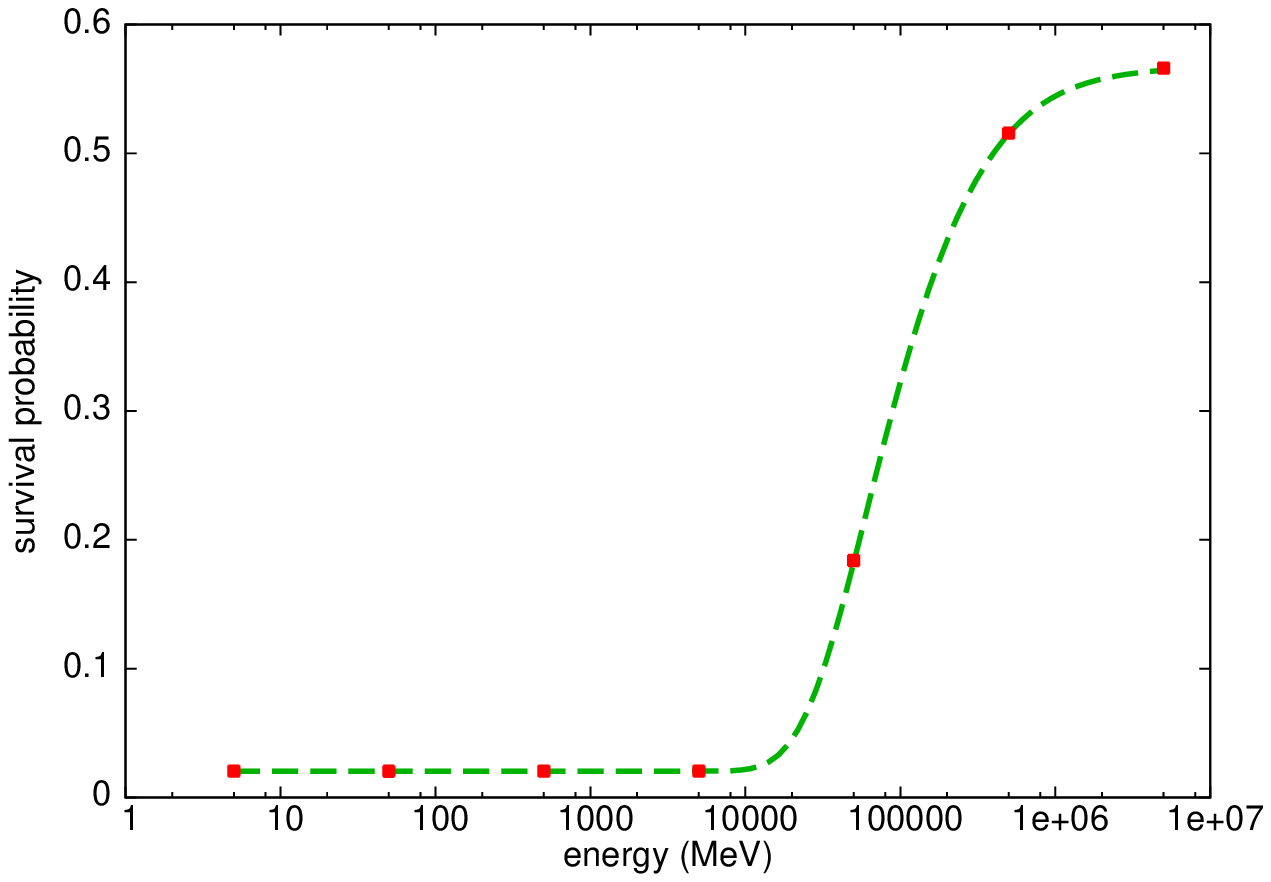}
\caption{The difference of $P(\nu_e \rightarrow \nu_e)$ between by the analytic solution assuming ICA (the broken line) and 
by the numerical calculation (the box points). Here, $\sin^22\theta_{13} = 0.08$ and 
$\Delta m^2_{31} = 2.4 \times 10^{-3} \, \mathrm{eV}^2$. } \label{fig:0.08}
\vspace*{0.5cm}
\end{center}
\end{figure}

\subsection{Energy dependence of ICA}
First of all, let us comment on the energy dependence of ICA. Since the 
Hamiltonian in vacuum is inversely proportional to the neutrino energy (see Eq.(\ref{eq:3flavor2})), the higher the neutrino 
energy is, the closer the distance of the two crossing points becomes. Figure \ref{fig:17} shows the asymptote of the energy diagram, in 
which the neutrino energy is $50 \mathrm{MeV}$. Although the density profile is of the exponential type, the higher point is 
so near to the lower point that the density profile looks like linear. We would like to discuss this problem in more detail. \\

In Fig.\ref{fig:0.08}, we have plotted the probability $P(\nu_e
\rightarrow \nu_e)$ by the analytic solution assuming ICA and that by the
numerical one for each neutrino energy.
The broken line (the box points) indicates the analytic (numerical) solution,
respectively. 
The set of the oscillation parameters
assumed here are
\begin{eqnarray*}
&& \sin^22\theta_{13} = 0.08 \\
&& \Delta m^2_{31} = 2.4 \times 10^{-3} \, \mathrm{eV}^2. 
\end{eqnarray*}
We assume the error of the numerical calculation within 0.005,
and we find no difference between the two solutions in Fig.\ref{fig:0.08}.
This result indicates that ICA holds even if the distance 
between two crossing points is close, i.e., the neutrino energy is high.

\begin{figure}[!hb]
\begin{center}
\includegraphics[width=10cm]{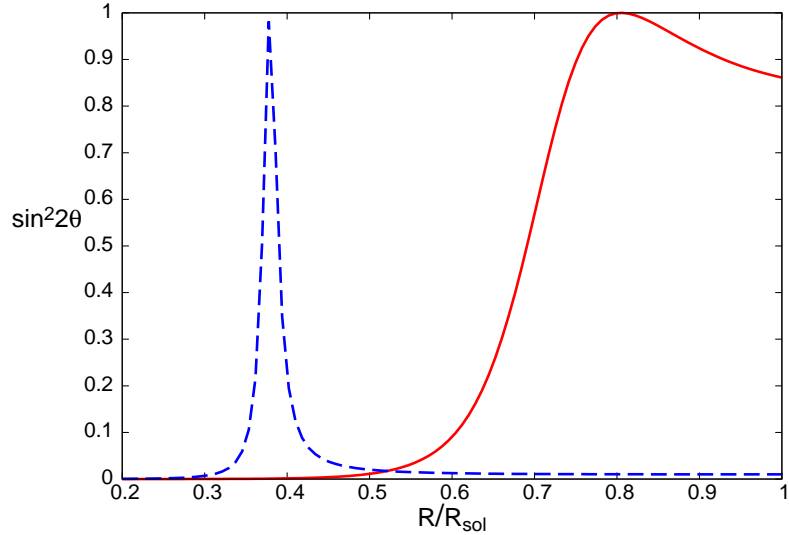}
\caption{Each shape of two resonances, non--overlapping. The left line is the higher 
resonance---Eq.(\ref{eq:appamp3}). The right line is the lower resonance---Eq.(\ref{eq:appamp2}).} \label{fig:19}
\end{center}
\end{figure}

From the analogy with Eq.(\ref{eq:appamp}), we get
\begin{subequations} \label{eq:appamp23}
\begin{align}
\sin^22\widetilde{\theta}_{12} &= \frac{(\Delta_{21}\sin2\theta_{12})^2}
			{(\Delta_{21}\cos2\theta_{12} - A)^2 + (\Delta_{21}\sin2\theta_{12})^2} \label{eq:appamp2} \\
\sin^22\widetilde{\theta}_{13} &= \frac{(\Delta\sin2\theta_{13})^2}
			{(\Delta\cos2\theta_{13} - A)^2 + (\Delta\sin2\theta_{13})^2}, \label{eq:appamp3}
\end{align}
\end{subequations}
where $\Delta$ has been defined in Eq.(\ref{eq:Delta}).
In Fig.\ref{fig:19}, we have plotted the shapes of Eq.(\ref{eq:appamp2})---the right line, and of Eq.(\ref{eq:appamp3})---the 
left line, in the case: 
\begin{eqnarray*}
\sin^22\theta_{13} &=& 1 \times 10^{-6} \\
\Delta m^2_{31} &=& 2.4 \times 10^{-3} \, \mathrm{eV}^2 \\
E &=& 700 \, \mathrm{MeV}, 
\end{eqnarray*}
as the neutrino propagate in the medium to vacuum. Thus, two resonances seem to be far apart. \\ \\

\newpage
Introducing the notations
\begin{eqnarray*}
y &\equiv& A/\Delta_{21} \\
\alpha &\equiv& \Delta/\Delta_{21} \, . 
\end{eqnarray*}
we can rewrite Eqs.(\ref{eq:appamp23}), 
\begin{subequations} \label{eq:appamp4}
\begin{align}
\sin^22\widetilde{\theta}_{12} &= \frac{\sin^22\theta_{12}}{(\cos2\theta_{12} - y)^2 + \sin^22\theta_{12}} \\
\sin^22\widetilde{\theta}_{13} &= \frac{(\alpha\sin2\theta_{13})^2}
							{(\alpha\cos2\theta_{13} - y)^2 + (\alpha\sin2\theta_{13})^2}. 
\end{align}
\end{subequations}
Thus, each resonance point is 
\begin{subequations} \label{eq:resp}
\begin{align}
y_H &= \alpha\cos2\theta_{13} \\
y_L &= \cos2\theta_{12} \, , 
\end{align}
\end{subequations}
and each half width at half maximums is 
\begin{subequations} \label{eq:hwhm}
\begin{align}
\Gamma_H &= \alpha\sin2\theta_{13} \\
\Gamma_L &= \sin2\theta_{12} \, . 
\end{align}
\end{subequations}
These quantities are independent of the neutrino energy. This means that two resonances never overlap even though 
the neutrino energy gets higher, i.e., even if the distance between two crossing points gets closer. 
From this one might be tempted to conclude that ICA always holds. 
The question we have to ask here is the validity of the 
non--overlapping resonance to judge whether ICA holds or not. \\

\begin{figure}[!h]
\begin{center}
\includegraphics[width=10cm]{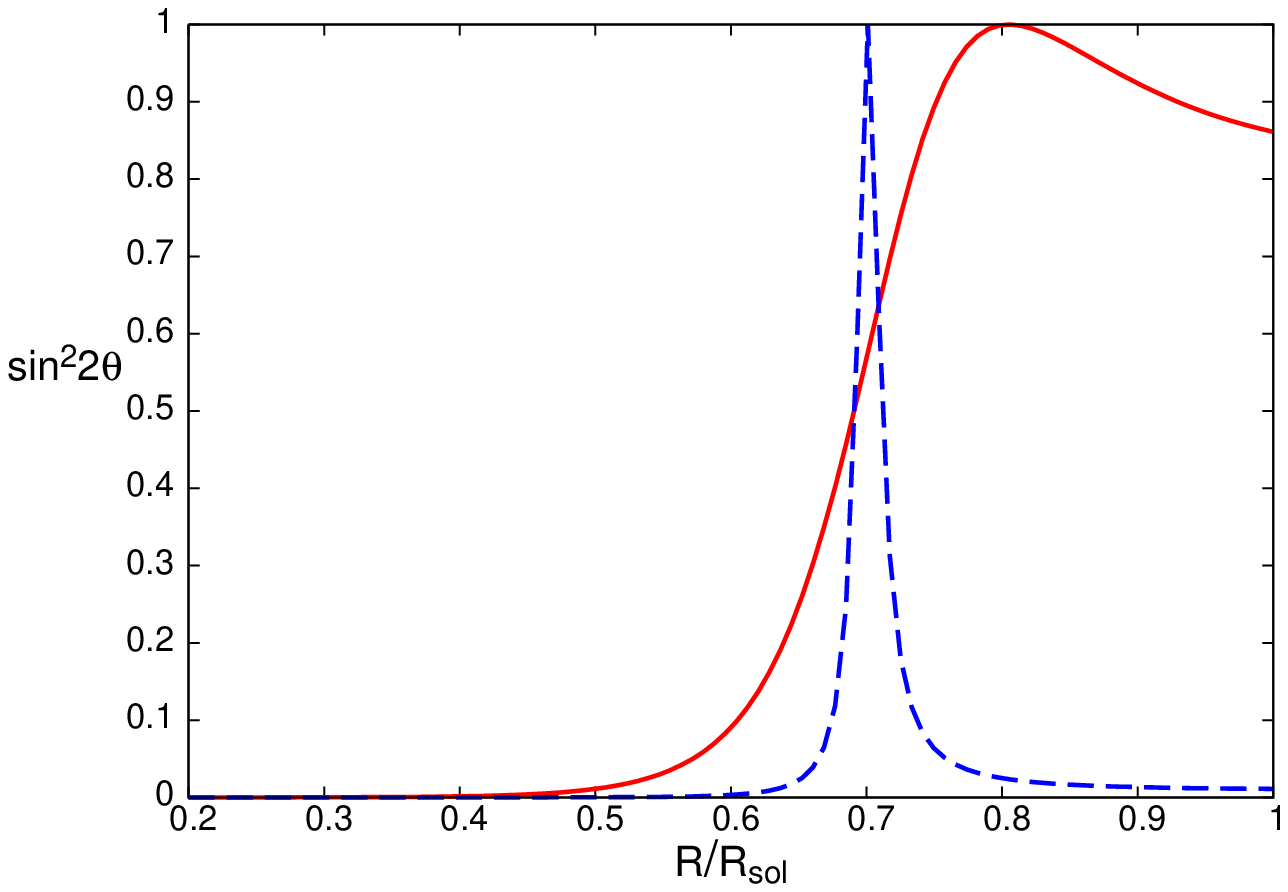}
\caption{The modified resonances---overlapping. Here, $\Delta m^2_{31}/\Delta m^2_{21} \simeq O(1)$. } \label{fig:20}
\vspace*{1.5cm}
\includegraphics[width=11cm]{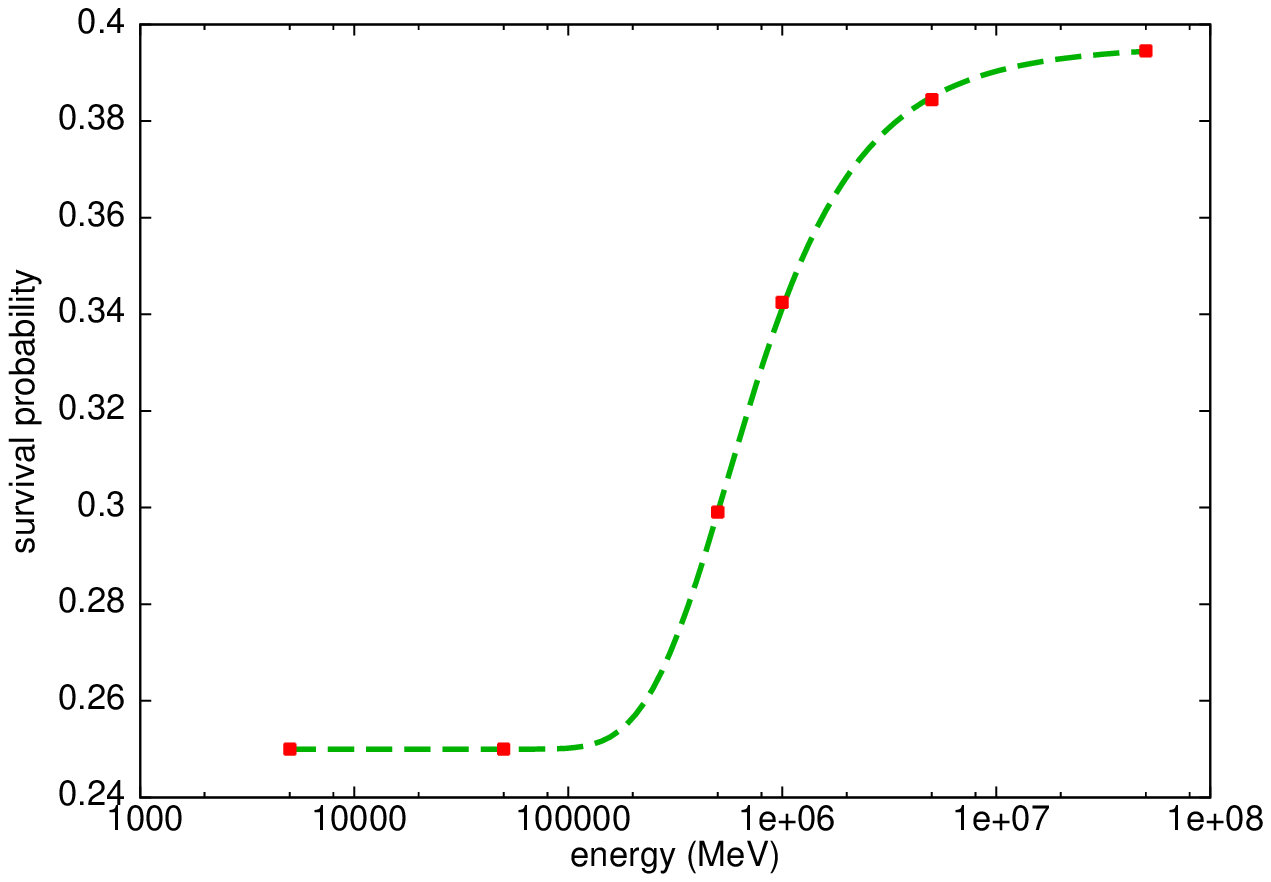}
\caption{The difference of $P(\nu_e \rightarrow \nu_e)$ between by the analytic solution assuming ICA (the broken line) and by the 
numerical calculation (the box points). Here, $\sin^22\theta_{13} = 0.75$ and 
$\Delta m^2_{31} = 2.4 \times 10^{-3} \, \mathrm{eV}^2$. } \label{fig:21}
\end{center}
\end{figure}

\subsection{ICA for overlapping resonances}
Let us now imagine the hypothetical situation: 
$$\Delta m^2_{31} \rightarrow 0.04 \times \Delta m^2_{31} \sim \Delta m^2_{21} \, , $$
that is, 
$$\Delta m^2_{31}/\Delta m^2_{21} \simeq O(1). $$
This parameter realizes the overlapping resonances (see Fig.\ref{fig:20}). Or, let us here imagine another hypothetical situation, 
$$\sin^22\theta_{13} = 0.75 \, (\theta_{13} = 30^\circ). $$
This parameter realizes the overlapping resonance too. 

In Fig.\ref{fig:21}, we have plotted the both probabilities $P(\nu_e \rightarrow \nu_e)$ for the second situation 
$(\sin^22\theta_{13} = 0.75)$. Again we find no difference between the two solutions.
From this we observe that overlapping of the two resonances is not
a sufficient condition for ICA to break down.

\subsection{Dependence of ICA on the mixing angle and the mass squared difference}
\begin{figure}[!h]
\begin{center}
\includegraphics[width=11cm]{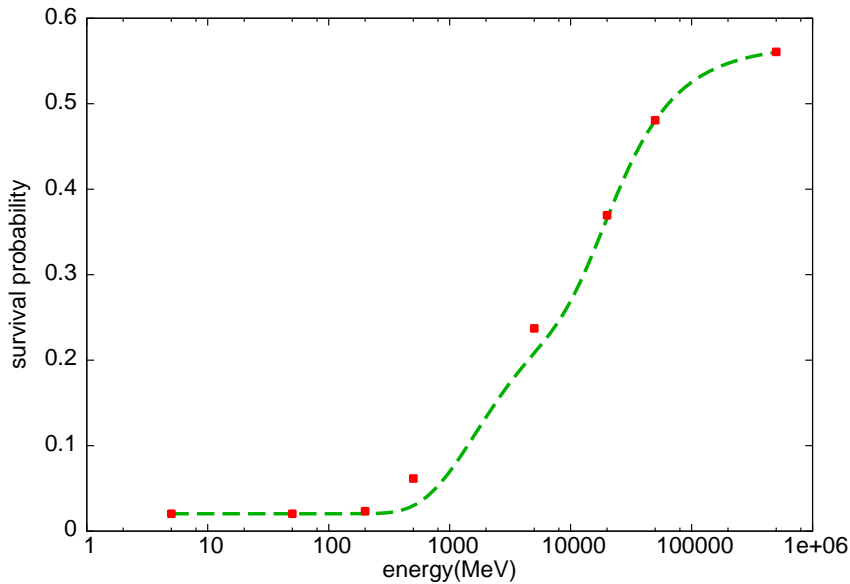}
\caption{The ICA breaking $(\Delta m^2_{31}/\Delta m^2_{21} \simeq O(1), \, \sin^22\theta_{13} = 0.08)$. } \label{fig:13_2}
\end{center}
\end{figure}

To search for the parameter range of ICA, we have examined several cases for $\theta_{13}$ and $\Delta m^2_{31}$, and tested 
whether ICA held or not. For example, in the case: 
\begin{eqnarray*}
\Delta m^2_{31}/\Delta m^2_{21} &\simeq& O(1) \\
\sin^22\theta_{13} &=& 0.08, 
\end{eqnarray*}
we found that
the ICA breaks down (see Fig.\ref{fig:13_2}). \\ \\

\begin{table}[!h]
\begin{center}
\caption{The difference of $P(\nu_e \rightarrow \nu_e)$ between by the analytic solution assuming ICA and by the numerical 
calculation. Nonzero terms mean that ICA is broken.} \label{tab:ica}
\includegraphics[width=4cm,angle=90]{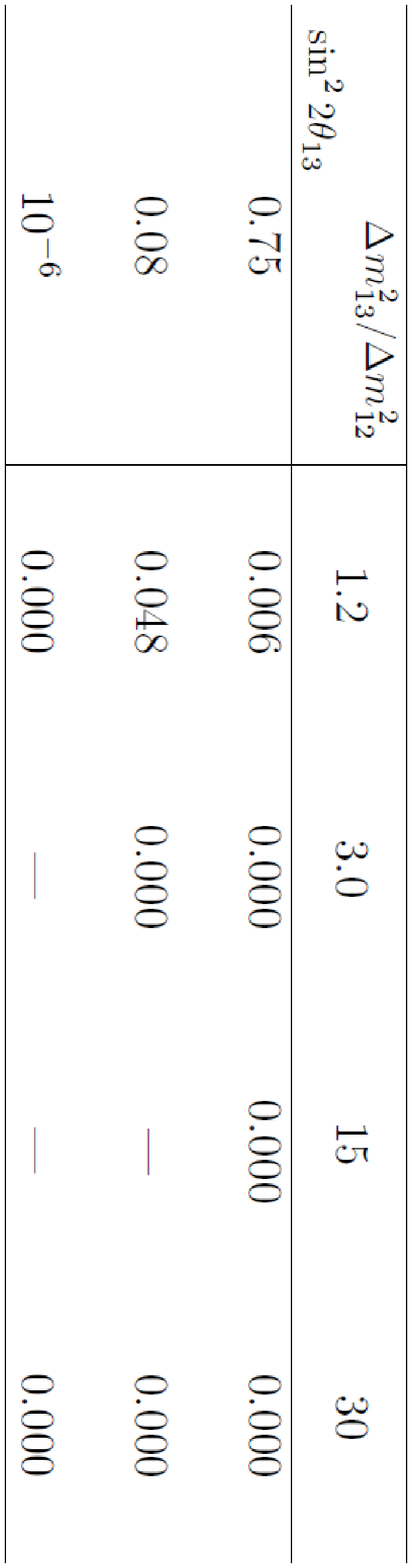}
\end{center}
\end{table}

\newpage
In Tab.\ref{tab:ica}, we have shown the difference between $P(\nu_e \rightarrow \nu_e)$ by the analytic solution assuming ICA 
and the one by the numerical one for several sets of $\sin^22\theta_{13}$ and $\Delta m^2_{31}/\Delta m^2_{21}$.  The 
numerical value 0.000 indicates that the difference is less than 0.005, the error of the numerical calculations. In such cases, we interpreted that ICA holds. 
The result implies that ICA is not applicable when $\theta_{13}$ is large and when $\Delta m^2_{31}/\Delta m^2_{21}$ is 
small.\footnote{This is equal to the condition for which the two flavor approximation fails.} \\

\subsection{The criterion of ICA}
Let us again consider the positive energy part of the Dirac equation for the neutrino in medium. From 
Eq.(\ref{eq:3flavor2}), we obtain
\begin{eqnarray} \label{eq:arb}
&& i\frac{d}{dt}\widetilde{\nu}_i = \widetilde{U}^{-1}H_f\widetilde{U}\widetilde{\nu}_i 
						- i\widetilde{U}^{-1}\frac{d\widetilde{U}}{dt}\widetilde{\nu}_i \, . 
\end{eqnarray}
In the adiabatic case, we can ignore the second term on the right hand side of Eq.(\ref{eq:arb}). Doing this matrix arithmetic, we 
have 
\begin{equation} \label{eq:3off}
i\frac{d}{dt}\left(
		  \begin{array}{c}
		  \widetilde{\nu}_1(t) \\ \\
	          \widetilde{\nu}_2(t) \\ \\
                  \widetilde{\nu}_3(t)
	          \end{array}
		\right)
 = \left(
     \begin{array}{ccc}
     \widetilde{E}_1 & -i\dot{\widetilde{\theta}}_{12} & -i\widetilde{c}_{12}\dot{\widetilde{\theta}}_{13} \\ && \\
     i\dot{\widetilde{\theta}}_{12} & \widetilde{E}_2 & i\widetilde{s}_{12}\dot{\widetilde{\theta}}_{13} \\ && \\
     i\widetilde{c}_{12}\dot{\widetilde{\theta}}_{13} & -i\widetilde{s}_{12}\dot{\widetilde{\theta}}_{13} & \widetilde{E}_3
     \end{array}
   \right)
          \left(
	    \begin{array}{c}
	    \widetilde{\nu}_1(t) \\ \\
	    \widetilde{\nu}_2(t) \\ \\
            \widetilde{\nu}_3(t)
            \end{array}
	  \right). 
\end{equation}
In order for ICA to apply to Eq.(\ref{eq:3off}), the Hamiltonian $H_L$ at the lower resonance has to be: 
\begin{subequations}
\begin{equation} \label{eq:12}
H_L = \left(
     \begin{array}{ccc}
     \widetilde{E}_1 & -i\dot{\widetilde{\theta}}_{12} & 0 \\
     i\dot{\widetilde{\theta}}_{12} & \widetilde{E}_2 & 0 \\
     0 & 0 & \widetilde{E}_3
     \end{array}
   \right). 
\end{equation}
Likewise, we have to have the following Hamiltonian $H_H$ at the higher resonance, for ICA to apply: 
\begin{equation}
H_H = \left(
     \begin{array}{ccc}
     \widetilde{E}_1 & 0 & 0 \\
     0 & \widetilde{E}_2 & i\dot{\widetilde{\theta}}_{13} \\
     0 & -i\dot{\widetilde{\theta}}_{13} & \widetilde{E}_3
     \end{array}
   \right). 
\end{equation}
\end{subequations}
In this case, we can get Eq.(\ref{eq:3level}).
From this, we observe that the ICA breaks down, for example, when the extra teams depending on $d\widetilde{\theta}_{13}/dt$ becomes large in $H_L$ at the lower resonance. 

Let us introduce the new parameter for the indicator of the ICA:
\begin{equation} \label{kl}
\kappa_L \equiv \left(\frac{d\widetilde{\theta}_{13}}{dt} \bigg/ \Delta\widetilde{E}_{21}\right)_L. 
\end{equation}
When this parameter is large, it means that the contribution of $d\widetilde{\theta}_{13}/dt$ is large in $H_L$. \\

In 
Fig.\ref{fig:g2}, where the x--axis (y--axis) is $\Delta m^2_{31}/\Delta m^2_{21}$
($\sin^22\theta_{13}$), we have drawn the contour lines of $\kappa_L$ with the 
neutrino energy satisfying $\gamma_L = 1$ (see Eq.(\ref{eq:ac})), for $\kappa_L = 1.0,\,0.5,\,0.1$. Thus, the new parameter 
$\kappa_L$ describes the results in Tab.\ref{tab:ica} very well. 

We can rewrite $\kappa^{-1}_L$ as
\begin{eqnarray*}
\kappa^{-1}_L &=& 
 \Delta_{21}\sin2\theta_{12} \bigg/ \left(\frac{\dot{A}\sin^22\widetilde{\theta}_{13}}{\Delta\sin2\theta_{13}}\right)_L 
 													      \nonumber \\
 &=& \frac{\Delta\sin2\theta_{12}}{\cos2\theta_{12}\displaystyle\left(\dot{n_e} / n_e\right)_L} \times
 	\frac{(\cos2\theta_{13} - \displaystyle(\Delta_{21} / \Delta)\cos2\theta_{12})^2 + \sin^22\theta_{13}}
        											{\sin2\theta_{13}}. 
\end{eqnarray*}
Fixing the neutrino energy by the condition $\gamma_L = 1$, we get 
\begin{subequations} \label{eq:klhin}
\begin{equation} \label{eq:klin}
\kappa^{-1}_L\,(E;\,\gamma_L = 1) = 
	\left(\frac{\Delta}{\Delta_{21}}\right)
        	\frac{(\cos2\theta_{13} - \displaystyle(\Delta_{21}/\Delta)\cos2\theta_{12})^2 + \sin^22\theta_{13}}
                								{\sin2\theta_{12}\sin2\theta_{13}}. 
\end{equation}

\begin{figure}[!h]
\begin{center}
\includegraphics[width=13.5cm,bb=60 94 400 250]{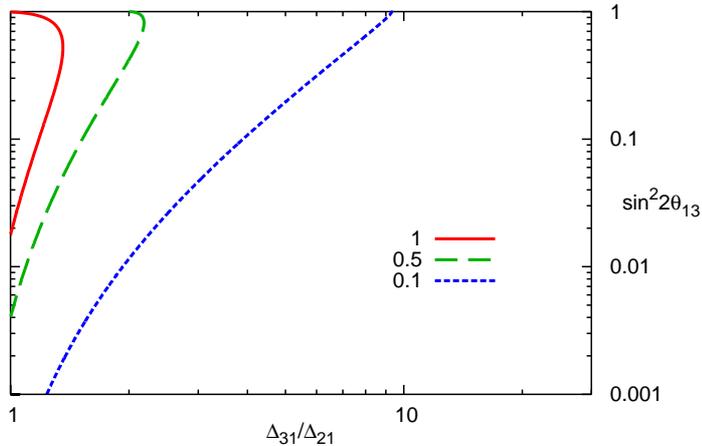}
\caption{The contour map of $\kappa_L$. } \label{fig:g2}
\end{center}
\end{figure}

As in Eq.(\ref{eq:klin}), we can get $\kappa^{-1}_H$ at the higher resonance
\begin{equation} \label{eq:khin}
\kappa^{-1}_H\,(E;\,\gamma_H = 1) = 
	\left(\frac{\Delta_{21}}{\Delta}\right)
        	\frac{(\displaystyle(\Delta/\Delta_{21})\cos2\theta_{13} - \cos2\theta_{12})^2 + \sin^22\theta_{12}}
                								{\sin2\theta_{12}\sin2\theta_{13}}. 
\end{equation}
\end{subequations}
Eqs.(\ref{eq:klhin}) are nothing but the condition of whether ICA holds or not, i.e., the 
condition for ICA as functions of $\theta_{13}$ and $\Delta m^2_{31}/\Delta m^2_{21}$.
It is remarkable that these 
parameters in Eqs.(\ref{eq:klhin}) are independent of the density profile $n_e(R)$. 

\section{Disccusion} \label{sec:dis}
In the previous section, we have investigated whether ICA held or
not, and the neutrino energy which requires ICA to break down ranges
in the interval $5\mathrm{MeV} \leq E \leq 50\mathrm{TeV}$.  At high
energy, the effect of the absorption of the neutrino by medium becomes
nonnegligible.
Let us consider this effect first.
The reaction number of the neutrino $N (/\mathrm{sec})$ is 
\begin{eqnarray*}
N &=& L \times \sigma(\nu N) \\
  &=& N_A \times \rho \times \sigma \times R \\
\sigma(\nu N)  &=& 0.68 \times 10^{-38} \times E \; (\mathrm{cm}^2), 
\end{eqnarray*}
where $L$ is the luminosity of neutrino, $\sigma(\nu N)$ is the cross section of neutrino--nucleus and $\rho$ is the matter 
density. Here we estimated the matter density $\rho = n_e(R)$ at $n_e(0) = \mathrm{const}$. Thus, the flight ranges at each 
neutrino energy for the present density profile Eq.(\ref{eq:density}) are 
\begin{eqnarray*}
R &\simeq& 2 \times 10^{-1} \:\, \mathrm{km} \quad \mathrm{(1PeV)} \\
  &\simeq& 2 \times 10^{2} \quad \mathrm{km} \quad \mathrm{(1TeV)} \\
  &\simeq& 2 \times 10^{5} \quad \mathrm{km} \quad \mathrm{(1GeV)}. 
\end{eqnarray*}
Although this is a rough estimate, we find that the neutrino, whose energy is not less than $1 \mathrm{TeV}$, cannot reach 
the vacuum area. Therefore, our hypothesis that high energy neutrinos are emitted
at the center of a supernova and are observed outside of the supernova may not make sense.

Secondly,
in dense matter for which there are two crossing points, the neutrino density $n_\nu(R)$ becomes large too. In such a case, the 
neutrino--neutrino interaction is 
strong~\cite{Duan:2010bg,Pantaleone:1992eq,Samuel:1993uw,Kostelecky:1994dt,Pastor:2001iu,Hannestad:2006nj,Fogli:2007bk}. 
Adding this effect, we need to incorporate $n_\nu(R)$, 
which is time--dependent, into the off--diagonal parts for $H$ (see Eqs.(\ref{eq:3flavors})). No formula of the 
nonadiabatic transitions attended by this effect have been established even the formulae with two flavor. 
In our study, we have assumed that the two jumping 
points are far apart from the range where
the neutrino--neutrino interactions become important.

\section{Conclusion}
In this paper, we addressed the question whether the Independent Crossing Approximation (ICA) holds or not.  Ref.~\cite{Kuo:1989qe}
has focused on the picture of the non--overlapping resonance by which they judge whether two nonadiabatic transitions are 
independent or not. One of the purposes of this study is to check whether this interpretation is correct or not. 

First of all, we have checked the dependence of ICA on the neutrino energy. Namely, we checked the dependence of ICA on the 
distance between the higher and lower crossing points.  By numerical calculations, we showed that ICA is 
independent of the neutrino energy, that is, the distance between two crossing points. 

Secondly, we have checked the validity of the picture of the non--overlapping resonance for ICA by varying the parameters 
$\theta_{13}$ and $\Delta m^2_{31}$ in the hypothetical ranges.  We found numerically that ICA can hold even if 
two resonances overlap, i.e., ICA does not always break down even if two resonant widths overlap. 

Thirdly, we have searched for the case in which ICA breaks down.  We found that ICA breaks down when $\theta_{13}$ is large and when $\Delta m^2_{31}/\Delta m^2_{21}$ is small. 

Finally, we have introduced the new parameters as the criterion of ICA. With this criterion of ICA, we interpret the ICA breaking as the sign of a large contribution of the
off--diagonal coefficients, for example, the contribution of $d\widetilde{\theta}_{13}/dt$ at the lower resonance.  We have shown that the new parameters---the criterion of ICA, taking the extra 
contribution into consideration, describe well whether ICA holds or not. 

From the recent experiments, we know that the ratio $\Delta m^2_{31}/\Delta
m^2_{21}$ is large~\cite{pdg:2008} and the mixing angle $\theta_{13}$
is small~\cite{Apollonio:2002gd}. Therefore the ICA breaking is tiny
enough for the two nonadiabatic transitions to be regarded as independent, and all
we need is just to multiply $\hat{P}_H$ and $\hat{P}_L$ which are
obtained by the two flavor formalism, as in Ref.~\cite{Kuo:1989qe}.

In this paper, we have dealt with the only the neutrino--electron
interaction. In other cases, for example, where neutrino transition moments couple to the
large magnetic fields~\cite{Cisneros:1970nq,Okun:1986na,Lim:1987tk,Akhmedov:1988uk},
under the influence of the neutrino--neutrino
interaction~\cite{Duan:2010bg,Pantaleone:1992eq,Samuel:1993uw,Kostelecky:1994dt,Pastor:2001iu,Hannestad:2006nj,Fogli:2007bk}, 
or under the influence of the
Non--Standard
Interaction~\cite{Wolfenstein:1977ue,Guzzo:1991hi,Roulet:1991sm}, our
procedure is necessary.

\subsubsection*{Ackowledgement}
The author received the great supports from Osamu Yasuda. I would like to thank Serguey Petcov, Takashi Aoki and Masumi 
Kanai for insightful comments and suggestions. 

The work was partially supported by the MEXT program "Support Program for Improving Graduate School Education". 


\end{document}